\documentclass[useAMS,usenatbib]{mn2e}
\usepackage{psfig}  
\pdfoutput=1

\usepackage{amstext}
\usepackage{amssymb}
\usepackage{amsmath, float}
\usepackage{graphicx}
\usepackage{txfonts}
\usepackage{amsmath,mathtools,revsymb,amssymb}
\usepackage{booktabs,multirow,calc,xspace,rotating}

\newcommand       \be           {\begin{equation}}
\newcommand       \ee           {\end{equation}}
\newcommand       \bea          {\begin{eqnarray}}
\newcommand       \eea          {\end{eqnarray}}
\newcommand       \kms		{\,{\rm km \,\, s}^{-1}}
\newcommand       \gcc		{\,{\rm g \,\, cm}^{-3}}

\newcommand       \mspy 	{\,{\rm M_\odot \, yr^{-1}}}

\newcommand       \M		{\,{\cal M_{\rm conv} }}

\newcommand       \ergs		{\,{\rm erg \,\, s}^{-1}}
\newcommand       \tnuc		{\rm t_{nuc}}
\newcommand       \Lc		{\rm L_{conv}}
\newcommand       \Lw		{\rm L_{wave}}
\newcommand       \Fw		{\rm F_{wave}}
\newcommand       \gl		{\rm \gamma_{leak}}
\newcommand       \gnu		{\rm \gamma_{\nu}}

\newcommand\plotone[1]
 {\centering \leavevmode \includegraphics[width={0.99\columnwidth}]{#1}}

\begin{document}

%%%%%%%%%%%%%%%%%%%%%%%%%%%%%%%%%%%%%%%%%%%%%%%%
\title[Mass Loss in the Last Year of Massive Stellar Evolution]{Wave-Driven Mass Loss in the Last Year of  Stellar Evolution:  Setting the Stage for the Most Luminous Core-Collapse Supernovae}
\author[E. Quataert \& J. Shiode]{E. Quataert$^{1}$\thanks{E-mail: eliot@berkeley.edu} \& J. Shiode$^{1}$\thanks{E-mail: jhshiode@berkeley.edu}  \\
  $^{1}$Astronomy Department and Theoretical Astrophysics
  Center, University of California, Berkeley, 601 Campbell Hall,
  Berkeley CA, 94720\\ }

%\date{Accepted . Received ; in original form }
%\pagerange{\pageref{firstpage}--\pageref{lastpage}} \pubyear{????}
\maketitle
%\label{firstpage}

\begin{abstract}
During the late stages of stellar evolution in massive stars (C fusion and later), the fusion  luminosity in the core of the star exceeds the star's Eddington luminosity.  This can drive vigorous convective motions which in turn excite internal gravity waves.   The local wave energy flux excited by convection is itself well above Eddington during the last few years in the life of the star.     We suggest that an interesting fraction of the energy in gravity waves can, in some cases, convert into sound waves as the gravity waves propagate (tunnel) towards the stellar surface.   The subsequent dissipation of the sound waves  can unbind up to several $M_\odot$ of the stellar envelope.   This wave-driven mass loss can explain the existence of extremely large stellar mass loss rates just prior to core-collapse, which are inferred via circumstellar interaction in some  core-collapse supernovae (e.g., SNe 2006gy and PTF 09uj, and even Type IIn supernovae more generally).  An outstanding question is understanding what stellar parameters (mass, rotation, metallicity, age) are the most susceptible to wave-driven mass loss.  This depends on the precise internal structure of massive stars and the power-spectrum of internal gravity waves excited by stellar convection.

\end{abstract}

\vspace{-0.7cm}
\section{Introduction}
\label{sec:int}
\voffset=-2cm
\vspace{-0.1cm}

%Some of the most extreme of these outbursts occur in the last year to decade of massive stellar evolution, when a small fraction of massive stars eject $\sim 1-10 \, M_\odot$ of their envelopes.

Many massive stars appear to lose a significant fraction of their mass in episodic outbursts rather than continuous line-driven winds (e.g., \citealt{bouret2005,smith2006}).  There is strong evidence from observations of  luminous supernovae (SNe) that the most extreme version of this phenomena is the ejection of $\sim 1-10 M_\odot$ of the stellar envelope  in the last year to decade of massive stellar evolution (in a very small fraction of massive stars).   In particular, the interaction between an outgoing supernova shock and such ejecta can explain some of the most optically luminous SNe yet detected (e.g., \citealt{smith2007b}), including, e.g., SN 2006gy \citep{smith2007,ofek2007} and perhaps the emerging class of hydrogen-poor ultraluminous SNe  \citep{quimby2011, chomiuk2011}.   In several cases, the late-time light curve disfavors one alternative explanation, that the luminosity is powered by unusually large amounts of radioactive Ni and Co  (e.g., \citealt{miller2010,chomiuk2011}).   Related evidence for prodigious mass loss in the last few years of stellar evolution comes from SNe like PTF 09uj, which was interpreted as shock-breakout from an extremely dense circumstellar wind \citep{ofek2010}.  Even  more typical Type IIn SNe (i.e., those displaying narrow emission lines indicative of circumstellar interaction) appear to require progenitor mass loss rates exceeding $\sim 10^{-2} \mspy$ \citep{kiewe2012}, far larger than can be explained by continuous mass loss processes operating in massive stars.

One of the central puzzles in the circumstellar interaction scenario for these SNe is why a massive star should lose a substantial fraction of its mass in only $\sim 10^{-6.5}$ of its lifetime.   This is required in order for the supernova shock to encounter the stellar ejecta at radii $\sim 100$ AU where the shock is particularly radiatively efficient. 

The most important change that occurs in the late stages of massive stellar evolution is the onset of prodigious neutrino cooling in the core of the star, associated with the high temperatures required for carbon fusion and beyond (see, e.g., \citealt{woosley02} for a review).    The fusion luminosities exceed Eddington for C fusion and later and become particularly large during the last year in the life of the star.   The rapid fusion and neutrino cooling in turn accelerate the nuclear burning so that the characteristic nuclear timescale $\tnuc$ (analogous to the main sequence timescale for H-burning) decreases to a $\sim$ yr for Ne and O fusion and a $\sim$ day for Si fusion (with the exact values depending on stellar mass, metallicity, and rotation).   Because neutrino cooling and fusion have different temperature sensitivities, it is in general not possible for neutrino cooling to balance nuclear heating everywhere in the core of the star, although in a volume averaged sense the two balance.   This local difference between heating and cooling drives convection which carries a significant convective luminosity $\sim 10s \,\%$ of the fusion luminosity.   

In this {\em Letter} we argue that wave excitation by  vigorous convection in the late stages of stellar evolution is capable of driving the strong mass loss suggested by circumstellar interaction in luminous core-collapse SNe.    Such wave excitation has been explicitly seen in  numerical simulations of carbon and oxygen shell fusion by Arnett and collaborators (e.g., \citealt{meakin2006}); there is also closely related numerical work in the context of solar convection (e.g.,  \citealt{rogers2005, rogers2006}).  Here we provide analytic estimates of  wave excitation in   evolved massive stars and discuss the resulting implications for mass loss in the last $\sim$ year of stellar evolution.  We begin by summarizing some of the key properties of convection during carbon fusion and later and the excitation of internal gravity waves by such convection (\S \ref{sec:conv}).  We then calculate the conditions under which a super-Eddington flux of waves excited in the core of the star can tunnel through to the stellar envelope (\S \ref{sec:damp}).   Finally, in \S \ref{sec:disc} we discuss the implications of our results and directions for future work.   

Because there are significant uncertainties in the interior structure of massive stars during the evolutionary phases of interest, we focus  on elucidating the general conditions required for efficient wave-driven mass loss.   We defer to future work the problem of finding stellar progenitors that have all of the requisite properties.   SNe with evidence for circumstellar interaction represent $\sim 10 \%$ of all core-collapse events \citep{smith2011}; ultraluminous SNe are much rarer still (e.g., \citealt{quimby2011}).   This suggests that rather special stellar parameters are required to generate $\sim M_\odot$ of ejecta in the last $\sim$ year of stellar evolution.  

\vspace{-0.7cm}
\section{Convection and Wave Excitation in Late Stages of Stellar Evolution}
\label{sec:conv}
\vspace{-0.1cm}

Table \ref{tab:conv} summarizes some of the key properties of core fusion and convection during the late stages of stellar evolution (based on \citealt{woosley02} and \citealt{kipp}).  
%The neutrino luminosities exceed Eddington for Carbon fusion and later and become particularly large during the last year in the life of the star.   The rapid neutrino cooling in turn accelerates the nuclear burning so that the characteristic nuclear timescale $\tnuc$ (analogous to the main sequence timescale for H-burning) decreases to a $\sim$ yr for Ne and O fusion and a $\sim$ day for Si fusion (with the exact values depending on stellar mass, metallicity, and rotation).  
%Because the temperature sensitivities of neutrino cooling and fusion are different, it is in general not possible for neutrino cooling to balance nuclear heating everywhere, although in a volume averaged sense the two balance.   This local difference between heating and cooling drives convection which  carries a luminosity $\Lc \sim L_\nu$.   
For a stellar core with a mass $\sim M_\odot$, a convective luminosity  $\Lc \sim L_{\rm fusion}$ corresponds to a typical convective velocity $v_c \sim 10 \, (\Lc/10^9 L_\odot)^{1/3} \, (\rho_c/10^7 \gcc)^{-1/9} \, \kms$, where $\rho_c$ is the central density.   The corresponding convective Mach numbers are given in Table \ref{tab:conv}, and are $\gtrsim 0.01$ for Ne fusion and later.   Although the parameters given in Table 1 are motivated by the core properties of evolved massive stars, similarly vigorous convection occurs during late stage shell burning.   For example, \citet{arnett2011} find convective luminosities and Mach numbers comparable to those given in Table \ref{tab:conv} for shell O fusion during the hour prior to core collapse.

\begin{table}
  \caption{Late Stages of Massive Stellar Evolution}
  \label{tab:conv}
  \begin{center}
    %% @{xxx} replaces spacing with xxx (used here to soak up space)
    \hspace*{-1cm}
    \begin{tabular}{ccccc}
      \toprule
       Stage & Duration \, ($\tnuc$) & ${\rm L_{fusion}} \, (L_\odot)$ & Mach ($\cal M_{\rm conv}$) & $\tau_c$ \, (s) \\
      \midrule
      \parbox{\widthof{Abell 1991 .}}{Carbon}  & \parbox{\widthof{Abell 4059}}{$\sim 10^3$ yr} & $\sim 10^{6} $  &  $\sim 0.003 $ & $\sim 10^{4.5}$ \\
      \parbox{\widthof{Abell 1991 .}}{Neon}  & \parbox{\widthof{Abell 4059}}{$\sim 1$ yr} & $\sim 10^{9}$ & $\sim 0.01$ & $\sim 10^3$ \\
      \parbox{\widthof{Abell 1991 .}}{Oxygen}  & \parbox{\widthof{Abell 4059}}{$\sim 1$ yr}  & $\sim 10^{10}$ & $\sim 0.02$ & $\sim 10^3$ \\
      \parbox{\widthof{Abell 1991 .}}{Silicon}  & \parbox{\widthof{Abell 4059}}{$\sim 1$ day} & $\sim 10^{12}$ & $\sim 0.05$ & $\sim 10^2$\\
      \bottomrule
    \end{tabular}
  \end{center}
  Fusion luminosities, durations,  convective Mach numbers, and convective turnover times for core fusion during the late stages of stellar evolution of a $\sim 25 M_\odot$ star (based on \citealt{woosley02}).   Precise values  depend somewhat on stellar mass, metallicity, and rotation.   Convective Mach number is an order of magnitude estimate assuming that a significant fraction of the fusion luminosity is locally carried by convection; depending on the stellar parameters, core carbon fusion may not be convectively unstable.   Shell fusion of C, O, Ne, etc. can produce similarly vigorous convection.
  \medskip
\end{table}

\begin{figure*}
\resizebox{14cm}{!}{\plotone{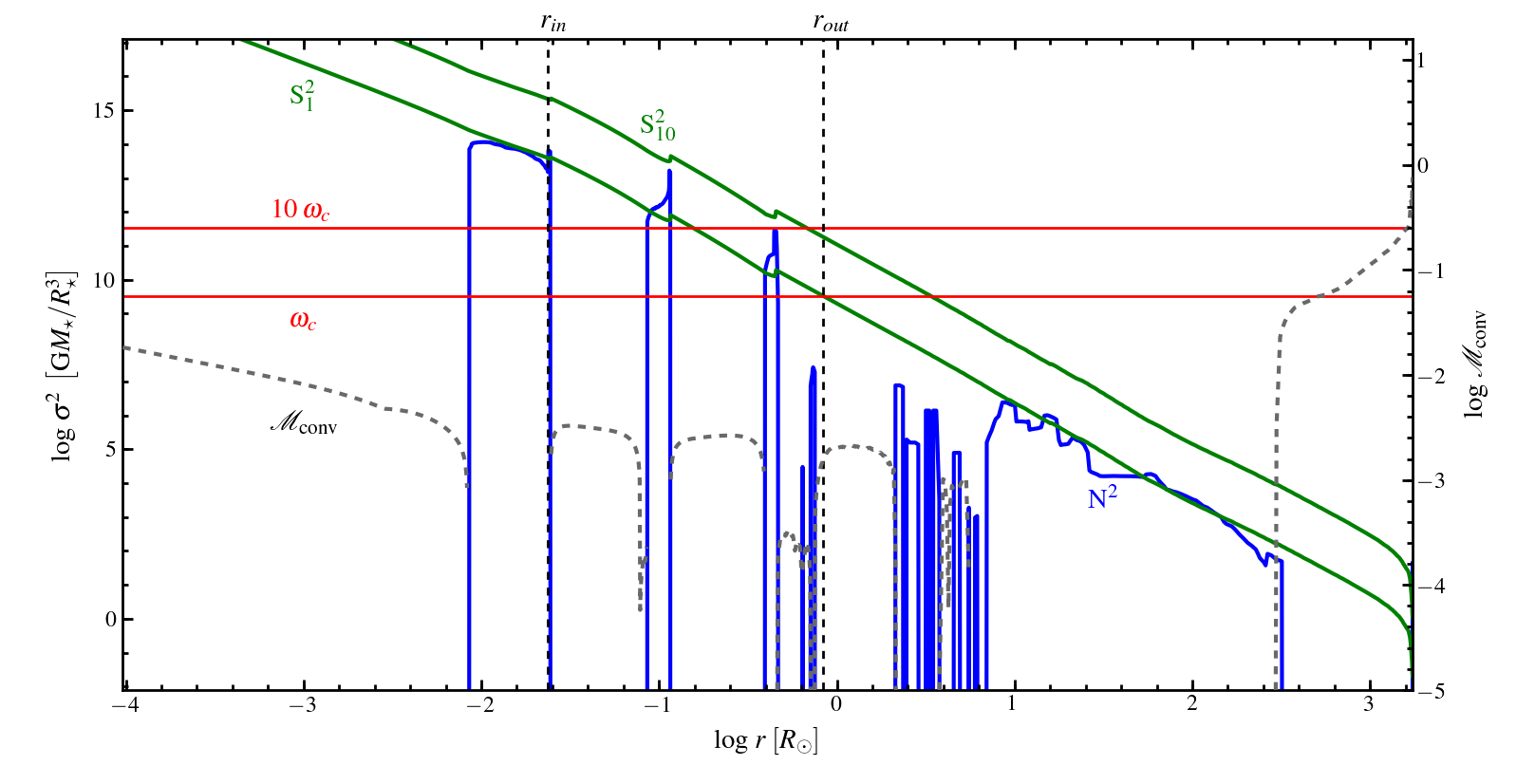}}
\caption{Propagation diagram for a 40 M$_\odot$ star with $Z = 10^{-4}$ during core O fusion, showing the Brunt-V\"ais\"al\"a frequency $N$ (blue, solid line) and the Lamb frequency $S_\ell$ (for $\ell = 1$ and 10; green, solid) on the left ordinate, plotted as a function of radius. The convective mach number ${\cal M}_{\rm conv}$ (gray, dashed) is also shown on the right ordinate. The thin red lines show the convective turnover frequency in the core, and ten times that frequency, to demonstrate the range of internal gravity wave frequencies the core convection is likely to excite. The inner and outer radius of the tunneling cavity ($r_{in}$ and $r_{out}$) for an $\ell = 1$ mode of frequency $\omega_{c}$ are marked on the top abcissa. 
}
\label{fig:prop}
\end{figure*}

Figure \ref{fig:prop} shows a mode propagation diagram for a 40 $M_\odot$, $Z = 10^{-4}$ metallicity model during core O fusion (at which point  $R \simeq 1700 \, R_\odot$, $T_{\rm eff} \simeq 4000$ K, and $L_{\rm photon} \simeq 10^{5.8} \, L_\odot$).   The model was evolved using the MESA 1D stellar evolution code \citep{mesa2011} with no mass loss. Convective boundaries are determined by the Schwarzschild criterion, and hydrogen- and non-burning convection zones have  overshooting of 1\% of the local pressure scale height.   This progenitor choice is somewhat arbitrary and thus unlikely to actually be the optimal progenitor for wave-driven mass loss.   We include this model to provide a quantitative example of the propagation diagram and likely mode properties during very late stages of stellar evolution.   For this particular stellar model, the core convection has a Mach number $\M \sim 0.01$ and locally carries a luminosity $\Lc \sim 0.1 \, L_{\rm fusion} \sim 10^{9.5} \, L_\odot$, with a comparable convective luminosity in the shell C fusion present at $r \sim 0.03 R_\odot$.

%%%%%%%%%%%%%%%%%%%%%%%%%%%%%%%%%%%%%% FIG 1 %%%%%%%%%%%%%%%%%%%%%%%%%%%%%%
%%%%%%%%%%%%%%%%%%%%%%%%%%%%%%%%%%%%%%%%%%%%%%%%%%%%%%%%%%%%%%%%%%%%%%%%%%%%

Vigorous convection transfers some fraction of the turbulent kinetic energy into waves, both sound waves and internal gravity waves.  For the specific stellar model in Fig. \ref{fig:prop}, the core convective region is bordered by a convectively stable region into which the internal gravity waves excited by convection can propagate.    This is important because the roughly incompressible convective motions are much more efficient at directly exciting internal gravity waves than acoustic waves  \citep{gk1990}.    Quantitatively, convection carrying a luminosity $\Lc$ with a Mach number $\cal M_{\rm conv}$ excites an internal gravity wave luminosity of \citep{gk1990}
\be
\Lw \sim {\cal M_{\rm conv}} \, \Lc \sim 10^{8} \left(\frac{\Lc}{10^{10} L_\odot}\right) \left(\frac{\cal M_{\rm conv}}{0.01}\right) \, L_\odot \label{lwave}.
\ee
This analytic estimate of the fraction of the convective energy transferred to internal gravity waves is  consistent (at the order of magnitude level) with simulations of internal gravity wave excitation by solar convection \citep{rogers2005,rogers2006}.  

The frequencies and wavelengths of the internal gravity waves are somewhat more uncertain than the total wave luminosity given in equation \ref{lwave}.   If the excitation is dominated by motions in the convection zone itself,  the characteristic frequency associated with most of the wave power is the convective turnover frequency, $\omega \sim \omega_c \equiv \tau_c^{-1}$.  Likewise, the horizontal spatial scale is set by the size of the convective eddies, so that the characteristic angular degree of the excited modes is $\ell \sim r/H$, where $H$ is the size of the convective eddies and $r$ is the radius where the excitation occurs (e.g., \citealt{kumar1999}).    For the core (and `thick' shell) convection relevant here, this would correspond to modes having $\ell \sim$ few.   

The uncertainty in these estimates is that much of the excitation may instead occur in a convective overshoot layer \citep{spruit1991,rogers2005,rogers2006}.  This increases both the frequency and angular degree $\ell$ of the excited waves because the length-scale over which the excitation happens is the thickness of the overshoot layer.   For evolved massive stellar models the thickness of the overshoot layer is particularly uncertain because the vigorous convection itself generates significant mixing and may substantially modify the structure of the star relative to that predicted by 1D models \citep{meakin2006,arnett2011}.   It is likely that the convective overshoot layer is significantly thicker in this circumstance than in the sun (where it is $\lesssim 0.05 H$ at the interior convective-radiative transition; e.g., \citealt{dalsgaard2011}).   This favors the excitation of lower $\ell$ waves with $\omega \sim \omega_c$.   For our analytic estimates we will use
\be
\omega \equiv \alpha \, \omega_c = \alpha \, {\cal M_{\rm conv}} \, \frac{c_s}{H} \sim \alpha  \, {\cal M_{\rm conv}} \, S_1
\label{om}
\ee
where $\alpha \sim 1-10$ is a dimensionless number that encapsulates the uncertainty associated with the wave excitation and $S_1$ is the $\ell = 1$ Lamb frequency.   The characteristic wave frequencies $\omega = \omega_c$ and $\omega = 10 \,\omega_c$ are shown with the thin red lines in Fig. \ref{fig:prop}.

\vspace{-0.5cm}
\section{Tunneling, Trapping, and Damping}
\label{sec:damp}

\subsection{Internal Gravity Waves}
\label{sec:grav}

Depending on their frequency and angular degree, the energy supplied to internal gravity waves can either remain trapped in the central part of the star or tunnel out to the stellar envelope in the form of  propagating sound waves.
%\footnote{Modes with both gravity wave and sound wave characteristics in different parts of a star are typically called "mixed modes" in the stellar seismology literature.}  
We now estimate the conditions under which the latter occurs.   

For internal gravity waves that become evanescent at a radius $r_{\rm in}$ and tunnel to a radius $r_{\rm out}$, the effective damping rate of the otherwise trapped mode due to leakage to large radii is given by \citep{unno} 
\be
\gl \sim \frac{v_g}{r_{in}} \, \left(\frac{r_{in}}{r_{out}}\right)^{2 \Lambda} \sim \omega \, \frac{\omega}{\Lambda \langle N \rangle} \, \left(\frac{r_{in}}{r_{out}}\right)^{2 \Lambda},
\label{gleak1}
\ee
where $\Lambda^2 = \ell\,(\ell+1)$ and $\langle N \rangle$ is the average Brunt-V\"ais\"al\"a frequency in the propagating region.   The radii $r_{\rm in}$ and $r_{\rm out}$ are labeled in Fig. \ref{fig:prop} for $\ell = 1$ modes with $\omega = \omega_c$.   Physically, the timescale for energy to leak out implied by equation \ref{gleak1} ($\sim \gl^{-1}$) is given by the group travel time across the propagating region ($r_{in}/v_g$) divided by the tunneling probability, i.e., the fraction of the mode energy that tunnels through the barrier in a given group travel time.    

In addition to the tunneling captured by equation \ref{gleak1}, the outgoing gravity waves can also be partially reflected by the rapid changes in composition at shell boundaries  (see the spikes in $N^2$ in Fig. \ref{fig:prop}); this occurs if the wavelength of the waves is larger than the thickness of the region over which the composition changes.   The latter is set by convective overshoot at the base of the shell convection zones and is probably a few percent of a scale height.   Given this, we estimate that lower frequency gravity waves with $\omega \sim \omega_c$ are reasonably in the WKB limit in which equation \ref{gleak1} applies, but for higher frequency waves with $\omega \sim 10 \omega_c$, the tunneling may be suppressed by an additional factor of $\sim 10$ due to the compositional boundaries.

Internal gravity waves of frequency $\omega$ (given by eq. \ref{om}) and degree $\ell$ excited at the outer edge of the core convection zone at radius $\sim r$ begin to tunnel at a radius $r_{\rm in} \sim 3 r$ (based on the width of the convectively stable region at $\sim 0.01 R_\odot$ in Fig. \ref{fig:prop}).   The radius where the tunneling ceases ($r_{\rm out}$) is determined by where $\omega > S_\ell$, so that the waves become propagating sound waves.   For massive stellar models at radii $\sim 10^{-2}-10 \, R_\odot$, we find that the Lamb frequency can be reasonably approximated as a power-law in radius $S_\ell \propto r^{-b}$, with $b \sim 1.2-1.5$.   Using equation \ref{om} for the characteristic wave frequency, this implies $r_{\rm out}/r_{\rm in} \simeq 0.3 \, (\alpha {\cal M_{\rm conv}})^{-1/b} \Lambda^{1/b}$ and thus
\be
\gl \sim  3^{2 \Lambda} \, \left(\alpha {\cal M_{\rm conv}} \Lambda^{-1} \right)^{2[\Lambda/b]+1} \, \omega.
\label{gleak}
\ee

The dominant  damping mechanisms for the gravity wave energy trapped in the core of the star are nonlinear damping and neutrino damping.   The neutrino damping rate is
\be
\gnu \sim 10 t_{th}^{-1} \sim 100 \, \tnuc^{-1}
\label{gnu}
\ee
where $t_{th}$ is the thermal (Kelvin-Helmholz) time and the factor of $10$ in front of $t_{th}^{-1}$ is due to the strong temperature dependence of the neutrino reactions (so that a given perturbation in temperature gives rise to a yet stronger perturbation in neutrino cooling).    
The last expression in equation \ref{gnu} is a consequence of the fact that the duration of nuclear burning ($\tnuc$ in Table \ref{tab:conv})  in the late stages of massive stellar evolution is only a factor of $\sim 10$ longer than the thermal time.   For some of the g-modes of interest it is possible that driving due to fusion may be stronger than the neutrino damping (see, e.g., \citealt{murphy2004}).  The interaction between convective excitation and such driving could be very interesting, but to be conservative, we do not explicitly include this driving in our estimates.

During O and Ne fusion, equation \ref{gnu} implies that $\gnu \sim 3 \times 10^{-6} \, {\rm Hz} \sim 10^{-3} \, \omega_c$, for $\omega_{c} \sim 3 \times 10^{-3}$ as shown in Fig. \ref{fig:prop}. Thus, neutrino damping is rather effective for modes of frequency $\sim \omega_{c}$.   For comparison, taking ${\cal M_{\rm conv}} \sim 0.01$, we find that $\gl \gtrsim \gnu$ for modes with $\ell = 1$, if $\alpha \gtrsim 3$.   For $\ell = 2$, the condition is  more prohibitive:   high frequency modes having $\alpha \gtrsim 10$ are required for leakage to dominate neutrino damping.   This demonstrates that only the power in the lowest degree modes with $\ell \simeq 1-2$ is capable of efficiently leaking out into the stellar envelope.   These low $\ell$ modes are, however,  those that are expected to be excited by the large-scale convection in late stages of stellar evolution. 

The nonlinear damping of  internal gravity waves is somewhat more difficult to quantify.   A useful measure of the nonlinearity in the propagating region is the dimensionless parameter $k_r \xi_r$ (the radial displacement relative to the radial wavelength):  when $k_r \xi_r \gtrsim 1$, the waves can locally overturn the stratification leading to efficient nonlinear damping.   Using conservation of energy flux, $k_r \xi_r$ for a traveling internal gravity wave of frequency $\omega$ can be rewritten as 
\be
k_r \xi_r \sim \Lambda^{3/2} \, \left( \frac{N}{\omega} \right)^{3/2} \left( \frac{\Fw}{\rho r^3 \omega^3} \right)^{1/2} \label{break}
\ee 
where $\Fw = \Lw /4 \pi r^2$ is the local wave energy flux.   Quantitatively, using the stellar progenitor shown in Fig. \ref{fig:prop}, and assuming $\Lw \sim 10^8 L_\odot$, we find that $k_r \xi_r \gtrsim 1$ for radii $\lesssim 0.015 \, R_\odot$ for $\omega \sim \omega_c$, while $k_r \xi_r \lesssim 0.01$ at all radii of interest for $\omega \sim 10 \omega_c$.   Wave breaking is thus unlikely to significantly limit the energies attained by higher frequency internal gravity waves, which also  tunnel the most effectively (this corresponds to larger $\alpha$ in eq. \ref{gleak}).   However, wave breaking may be important for waves with frequencies $\sim \omega_c$.  Indeed, \citet{meakin2006} see some evidence for mixing induced by g-modes breaking in the convectively stable region separating O and C-burning shells in their simulations of late-stage burning in massive stars.

\vspace{-0.3cm}
\subsection{Sound Waves}
\label{sec:sound}

%%%%%%%%%%%%%%%%%%%%%%%%%%%%%%%%%%%%%% FIG 2 %%%%%%%%%%%%%%%%%%%%%%%%%%%%%%
\begin{figure}
\plotone{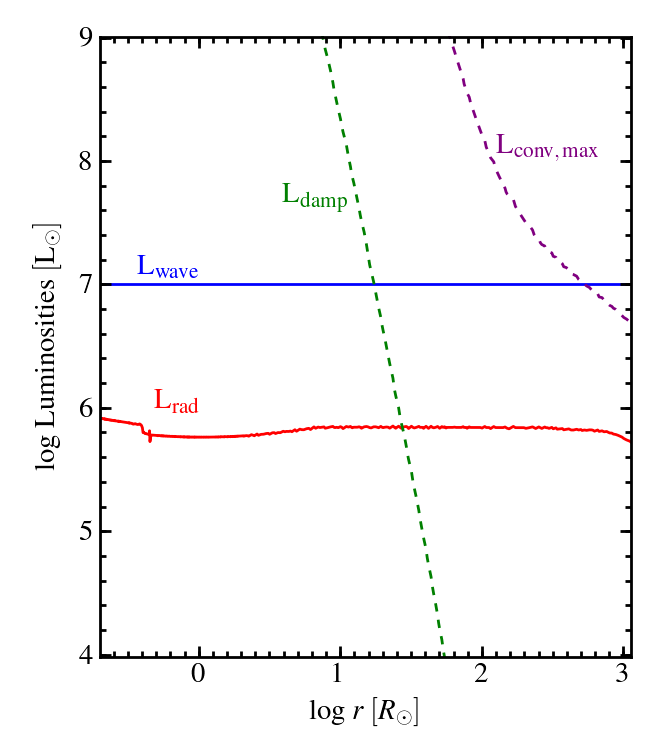}
\caption{Critical luminosities that determine the dissipation of the outgoing wave power $L_{\rm wave}$ and the ability of convection to carry the thermalized energy.   All quantities are for the $40 M_\odot$, $Z = 10^{-4}$ stellar model from Fig. \ref{fig:prop}.   Outgoing acoustic waves damp by radiative diffusion when the background radiative flux in the stellar model $L_{\rm rad}$ exceeds the critical luminosity $L_{\rm damp}$ (see eq. \ref{lcrit}; $\omega = 3\, \omega_c$ for $L_{\rm damp}$ in this example).  Subsonic convection cannot carry the thermalized wave power ($L_{\rm wave} = 10^7 L_\odot$ here) when $L_{\rm wave} \gtrsim L_{\rm conv, max} = 4 \pi r^2 \rho c_s^3$.  Taken together, these results imply that the super-Eddington wave power cannot be carried by either convection or radiative diffusion in the stellar envelope at $r \gtrsim 500 R_\odot$,  leading to strong mass loss.  There is $\sim 5 M_\odot$ exterior to this radius in this example.}
\label{fig:lums}
\end{figure}
%%%%%%%%%%%%%%%%%%%%%%%%%%%%%%%%%%%%%%%%%%%%%%%%%%%%%%%%%%%%%%%%%%%%%%%%%%%%

The outward energy flux in sound waves created as described in \S \ref{sec:grav} dissipates well before reaching the stellar surface.   In particular, we find that the dominant dissipation of the sound waves is via radiative diffusion.   The sound waves damp when the radiative diffusion time $\sim 1/(k^2 \chi)$ (where $\chi$ is the radiative conductivity) is shorter than the group travel time of the modes across a given scale-height of the star $\sim H/c_s$.   This condition can be rewritten as 
\be
L_{\rm rad} \gtrsim L_{\rm damp} \equiv \frac{4 \pi r^2 \rho c_s^3}{(k H)^2},
% \left(\frac{p_{\rm tot}}{p_{\rm gas}}\right).
\label{lcrit}
\ee  
where we have assumed that gas pressure is not much smaller than radiation pressure (which is reasonable for the conditions of interest).  Fig. \ref{fig:lums} shows $L_{\rm rad}$ (which is $\sim L_{\rm Edd}$ even though the envelope is convective) and $L_{\rm damp}$ as a function of radius for modes with $\omega \sim 3\,\omega_c$ in the 40 $M_\odot$, $Z = 10^{-4}$ progenitor used in Fig. \ref{fig:prop}.   The sound waves damp by radiative diffusion in the envelope at $r \sim 30-100 \, R_\odot$.  

The dissipation of the outgoing sound waves will drive convection in the  outer stellar envelope.   However, the convection ceases to be efficient when the wave luminosity is larger than the maximum energy that subsonic convection can carry $L_{\rm max, conv} = 4 \pi r^2 \rho \,c_s^3$.    So long as $\Lw \gtrsim L_{\rm Edd}$, this will always occur  inside the photosphere.  For the specific model shown in Fig. \ref{fig:lums}, $\Lw > L_{\rm conv, max}$ outside $\sim 500 R_\odot$.  There is $\sim 5 M_\odot$ of mass exterior to this radius.    The inability of convection or radiative diffusion to carry the outgoing wave energy implies that the wave power will inevitably drive a strong outflow.  In particular,  the mass outflow rate induced is likely to be $\dot M \sim 4 \pi r^2 \rho c_s$ (evaluated where $\Lw \sim L_{\rm conv, max}$, so that convection cannot carry the energy; this corresponds to the sonic point of the outflow).   For the stellar model used in Figures \ref{fig:prop} and \ref{fig:lums},  we find $\dot M \sim 10 \mspy$ and $\Lw \gtrsim \dot M v_{esc}^2$ (for $\Lw \gtrsim 10^{7} \, L_\odot$), so that the wave power is capable of driving a sustained outflow.

\vspace{-0.7cm}
\section{Discussion}
\label{sec:disc}
\vspace{-0.15cm}

The total energy released during Ne and O fusion in the last year of massive stellar evolution is $\sim 10^{51}$ ergs.  We have shown that a significant fraction of this energy $\sim 10^{48-49}$ ergs -- i.e., $\Lw \sim 10^{41-42} \, \ergs$ -- is converted into internal gravity waves via the Mach $\gtrsim 0.01$ convection that accompanies the enormous nuclear and neutrino luminosities in the cores of massive stars.   We have further argued that a large fraction of the power in low angular degree  ($\ell \sim$ few)  modes can tunnel through to the stellar envelope and become outgoing acoustic waves.   

If the low degree modes carry a significant fraction of the total wave power, the outgoing energy flux in acoustic waves will be significantly super-Eddington.   The dissipation of the acoustic waves in the stellar envelope then drives convection  that attempts to carry a  super-Eddington power $\sim 10^{40-41} \ergs \sim 10-100 \, L_{\rm Edd}$.   Such convection becomes inefficient  inside the stellar photosphere (Fig. \ref{fig:lums}) and thus the end result of this wave energy deposition is almost certainly substantial mass loss.    Assuming that the outgoing wave power can be maintained for a reasonable fraction of the duration of Ne and O core fusion, up to $10^{47-48}$ ergs is deposited in the stellar envelope.  If the unbound material moves at $\sim 100-1000 \kms$, the wave power can unbind $\sim 10^{-2}-10 \,M_\odot$ of material.   

Matter moving at $\sim 300 \kms$ ejected in the year prior to core-collapse will end up at $\sim 100$ AU when the star explodes.   The mass-loss rates, ejecta mass, and the radial extent of the ejecta estimated here are consistent with the conditions required to explain SNe  powered by circumstellar interaction between the SN shock and surrounding stellar ejecta.  In particular, at the low end of our estimated mass loss rates ($\sim 10^{-3}-10^{-2} \mspy$),  circumstellar interaction can produce typical Type IIn supernovae \citep{kiewe2012}.  More extreme cases with ejecta masses approaching $\sim 0.1-1 M_\odot$ in the last year are comparable to what is needed to power  the most  luminous SNe ever detected (e.g., SN 2006gy, \citealt{smith2007}; and the ultraluminous Ics, \citealt{quimby2011}) as well as SNe powered by shock breakout in a dense circumstellar wind (e.g., \citealt{ofek2010}).   Moreover, our results provide an explanation for what is otherwise a  fine tuning problem:  why should the star happen to lose a non-negligible fraction of its mass in the last year of stellar evolution ($\sim 10^{-6.5}$ of its lifetime!)?

For red super-giant progenitors, the mechanism of mass loss proposed here operates primarily during core Ne and O fusion.   During core C fusion, the energy flux in internal gravity waves is typically well below Eddington and so is unlikely to modify the stellar mass loss.   During Si fusion, the wave luminosities are  enormous, but the sound crossing time of a giant ($\sim$ few months) is much longer than the duration of the burning phase and so the star undergoes core-collapse before waves can  reach the stellar surface.   Nonetheless, the outgoing wave power created during Si fusion may substantially modify the structure of the progenitor star at radii $\lesssim 30 R_\odot$ \citep[see][]{meakin2006}.   This could change how the supernova shock interacts with the surrounding star.   In addition, in compact progenitors that have already lost their hydrogen envelopes, wave excitation during Si fusion could help drive substantial mass loss in the day prior to core-collapse.   Such stripped envelope progenitors are of particular interest in the context of understanding whether circumstellar interaction powers hydrogen-poor ultra-luminous SNe  (e.g., \citealt{quimby2011}).

In the well-studied case of SN 2006gy -- whose lightcurve is consistent with $\sim 10 M_\odot$ of ejecta in the 8 years prior to core-collapse \citep{smith2010} -- there is also evidence for a comparable amount of ejecta $\sim 10^3$ years earlier (via an IR light echo; \citealt{miller2010}).   The wave-driven mass loss mechanism we have proposed cannot work in its present form for this earlier mass loss episode.   Likewise, it is unlikely to be relevant to typical luminous-blue variable outbursts.  

Our model for wave-driven mass loss requires that a reasonable fraction of the internal gravity wave power excited by convection reside in low $\ell$ modes with frequencies somewhat larger than the characteristic convective turnover frequency.  The reason for the former condition is that  high $\ell$ gravity waves cannot tunnel through to the stellar surface; their energy is instead trapped in the interior, where it is ultimately thermalized and radiated via neutrinos.  The latter condition is required since lower frequency gravity waves will likely break locally within the g-mode propagating region and induce mixing (see eq. \ref{break}; such mixing could be very interesting in its own right:  e.g., by bringing fresh fuel down to higher temperatures where it could in principle combust and power an eruption like those studied by \citealt{dessart2010}).  In their simulations of O shell fusion, \citet{meakin2006} found significant wave power  in $\ell = 4$ modes; since these simulations covered only a quadrant of the star, these were the lowest $\ell$ modes they could simulate.   This is consistent with our hypothesis that the  convection in the late stages of stellar evolution will be particularly large-scale and thus will excite low $\ell$ modes.   

It is unclear whether core convection or shell convection is likely to be the most important source of waves for driving mass loss.   Core convection  tends to be more energetic and is likely larger scale, favoring the excitation of the low $\ell$ modes that tunnel most effectively.   However,  a countervailing consideration is that waves excited during shell convection have a smaller barrier to tunnel through.   Further numerical work to calibrate the power-spectrum of wave excitation is clearly needed, particularly during core O and Ne fusion.   

The basic energetics of core convection and wave excitation that we have highlighted apply to all massive stars.    It is, however, clear on observational grounds that not all massive stars undergo extreme mass loss in the year prior to core-collapse.    Given that our mechanism requires that a significant fraction of the gravity wave power excited by convection is in relatively low $\ell$ modes and that the tunneling cavity not be too spatially extended, we suspect that the answer to this apparent tension is that only in certain progenitors (with particular mass, metallicity, and/or rotation) are these conditions satisfied.   However, determining exactly which progenitors these are will require additional work.   For example, as emphasized by \citet{arnett2011}, the structure of the core of the star itself depends on the mixing induced by the internal gravity waves and so standard one-dimensional models may not be sufficiently accurate to address these questions.    

A second question that we have not addressed is how the star responds to the wave power deposited in the stellar envelope.   It is possible that in some cases the stellar structure adjusts in such a way as to suppress the fraction of the wave power that can tunnel to large radii before much of the stellar envelope has been shed.   This would limit the total amount of mass loss via wave energy deposition.    This question will be explored in detail in future work.

\vspace{-0.5cm}
\section*{Acknowledgments}
\vspace{-0.1cm}
We are grateful to Nathan Smith for many stimulating discussions that inspired us to think about this problem.  We thank Bill Paxton for  useful exchanges about MESA and  Avishay Gal-Yam, Chris Kochanek, and the referee for useful comments.  EQ was supported in part by the David and Lucile Packard Foundation.  This work was also supported by NASA Headquarters under the NASA Earth and Space Science Fellowship Program Grant 10-Astro10F-0030.

  \vspace{-0.5cm}
%%%%%%%%%%%%%%%%%%%%%%%%%%%%%%%%%%%%%%%%%%%%%%%%%%%%%%%%%%%%%%%%%%
\bibliography{refs}

%\label{lastpage}

\end{document}